\documentclass[fleqn,usenatbib]{mnras}

\usepackage{newtxtext,newtxmath}

\usepackage[T1]{fontenc}

\DeclareRobustCommand{\VAN}[3]{#2}
\let\VANthebibliography\thebibliography
\def\thebibliography{\DeclareRobustCommand{\VAN}[3]{##3}\VANthebibliography}

\newcommand{\appropto}{\mathrel{\vcenter{
  \offinterlineskip\halign{\hfil$##$\cr
    \propto\cr\noalign{\kern2pt}\sim\cr\noalign{\kern-2pt}}}}}
\usepackage{colortbl}
\usepackage{multirow}
\usepackage{booktabs}

\usepackage{xcolor}   

\usepackage{graphicx}	
\usepackage{amsmath}	

\definecolor{gbox}{HTML}{a3cef1}

\definecolor{grbl}{rgb}{0.3,0.6,0.7}

\title[Intrinsic galaxy alignments in CAMELS]{Intrinsic galaxy alignments in CAMELS simulations and the significant impact of baryon model}

\author[D. Bilsborrow, N. Jeffrey]{
Daniel J. L. Bilsborrow$^{1}$\thanks{E-mail: daniel.bilsborrow.20@ucl.ac.uk} and 
Niall Jeffrey$^{1}$\thanks{E-mail: n.jeffrey@ucl.ac.uk}
\\
$^{1}$Department of Physics and Astronomy, University College London, London WC1E 6BT, United Kingdom
}

\date{Accepted XXX. Received YYY; in original form ZZZ}

\pubyear{\the\year{}}

\begin{document}
\label{firstpage}
\pagerange{\pageref{firstpage}--\pageref{lastpage}}
\maketitle

\begin{abstract}
We present a detection of the intrinsic galaxy alignments in the CAMELS suite of hydrodynamic simulations. We find that the alignment amplitude depends significantly on cosmological and supernova feedback parameters—specifically $\Omega_m$, $\sigma_8$, $A_{\text{SN1}}$, $A_{\text{SN2}}$—while no dependence on AGN feedback is observed (due to the limited simulation volume $(25\,h^{-1}\,\text{Mpc})^3$). The dependence on $\sigma_8$ vanishes when projected correlation functions $w_{m+}$ are normalized by matter density correlations $w_{mm}$, consistent with predictions from linear alignment models. We find alignment amplitudes in quiescent galaxies to exceed those in star-forming galaxies by an order of magnitude. Moreover, examining orientation-only correlation functions from ellipticity-normalized galaxies $\tilde w_{m+}$, we confirm that alignment signals retain sensitivity to supernova feedback across full, star forming, quiescent, and ellipticity-normalized samples. Finally, we find evidence that supernova feedback impacts alignment signals differently in star-forming versus quiescent populations, suggesting that distinct alignment mechanisms operate across galaxy types. Our results offer key insights for understanding galaxy formation and alignment models for future weak gravitational lensing analyses.
\end{abstract}



\section{Introduction}
The intrinsic alignments of galaxies act both as a novel probe of galaxy formation and evolution \citep[e.g.][]{time_evolution_ia, Pk_TNG, sami_bh} and as a significant source of systematic error in {weak gravitational lensing analyses \citep[e.g.][]{Hitara2004, krause_ia, amon, Secco_2022, KiDS25}.} Accurate modelling of intrinsic alignments is therefore essential – not only to extract cosmological information reliably from lensing surveys, but also to deepen our understanding of the processes affecting galaxy evolution.

However, uncertainty in intrinsic alignment modelling remains a bottleneck for cosmological inference from state-of-the-art photometric {galaxy surveys \citep{Kirk_2010, Joachimi_2015, Paopiamsap_2024}.} It is, therefore, necessary to understand the mechanism of intrinsic alignments for different populations of galaxies under varying cosmological and astrophysical model assumptions. A further major source of uncertainty in weak lensing modelling are baryonic feedback processes, which also alter the matter distribution and impact observed signals. Measuring the relationship between intrinsic alignments and baryonic physics in simulations is a step towards mitigating modelling biases and, possibly, improving the containing power of cosmological datasets.

Understanding the alignment mechanism is particularly important for realistic semi-empirical prescriptions, necessary for pipeline validation \citep{lange_desi}, field-level likelihood inference \citep{borg}, and simulation-based inference \citep{2025A&A...694A.223V, sbi}. But, even when using classical (less-constraining) techniques for cosmological inference, mismatches between assumed baryonic feedback models and intrinsic alignment prescriptions can introduce significant biases in cosmological parameter estimation, especially if they are not consistent with the properties of the particular galaxy sample.


{
There are well established theoretical frameworks describing intrinsic alignments in different galaxy populations. For early-type, quiescent galaxies, observations demonstrate significant correlations between galaxy shapes and the density field \citep{Singh_2015, 2019Johnston, 2021A&A...654A..76F}, broadly consistent with tidal-alignment models where stellar distributions are elongated along the direction of surrounding overdensities \citep{catelan2001, Hitara2004, Blazek_2015}. In contrast, late-type or star-forming galaxies are expected to align primarily through angular momentum acquisition: tidal torques during halo formation generate galaxy spin directions that are tangentially correlated with the surrounding tidal field \citep{Doroshkevich1970, 1984ApJ...286...38W}. Observational studies have found evidence for correlated spin axes in spiral galaxies \citep{Pen_2000, Mandelbaum2010, Jones2010, chisari_15}, although corresponding two-point correlations of projected galaxy shapes are significantly weaker \citep{mandelbaum2011, 2019Johnston}, reflecting the lower amplitude of tidal-torque-induced intrinsic alignments. Motivated by early observations, \citet{Crittenden_2001} introduced a quadratic alignment model for spiral galaxies, modelling systems as thin discs whose observed ellipticities arise from angular-momentum-induced couplings to the large-scale tidal field. While \citet{Blazek_2019} proposed a perturbative approach that includes both the tidal shearing and tidal torquing alignment mechanism. Recent work \citep{Joachimi_2015, Barsanti_2022, 10.1093/mnras/stad2728} suggests that baryonic feedback, mergers, and morphological evolution may blur the distinction between alignment mechanisms in early- and late-type galaxies.}

In this work, we have both measured the intrinsic shapes of galaxies across 1000 CAMELS simulations~\citep{Villaescusa_Navarro_2021} and detected their alignment with large-scale structure. By cross-correlating the projected galaxy ellipticities with the underlying dark matter, we sidestep considerations of galaxy bias to obtain a clean two-point alignment signal.

Previous analyses have measured intrinsic alignments in simulations, but with fixed (or restricted) cosmological and astrophysical simulation parameter values \citep[e.g][]{2015SSRv..193...67K, chisari_15, 2020Zjupa, 2021MNRAS.501.5859T, samuroff_sim, Delgado_2023}.

The variations of cosmological and astrophysical parameters used in CAMELS allow us – for the first time – to directly measure the dependence of the intrinsic galaxy alignments on these parameters over a wide, realistic range. We additionally explore the alignment mechanism by analysing subsamples of galaxy type and also by measuring orientation-only correlations.

\begin{figure*}
	\includegraphics[width=1.0\linewidth]{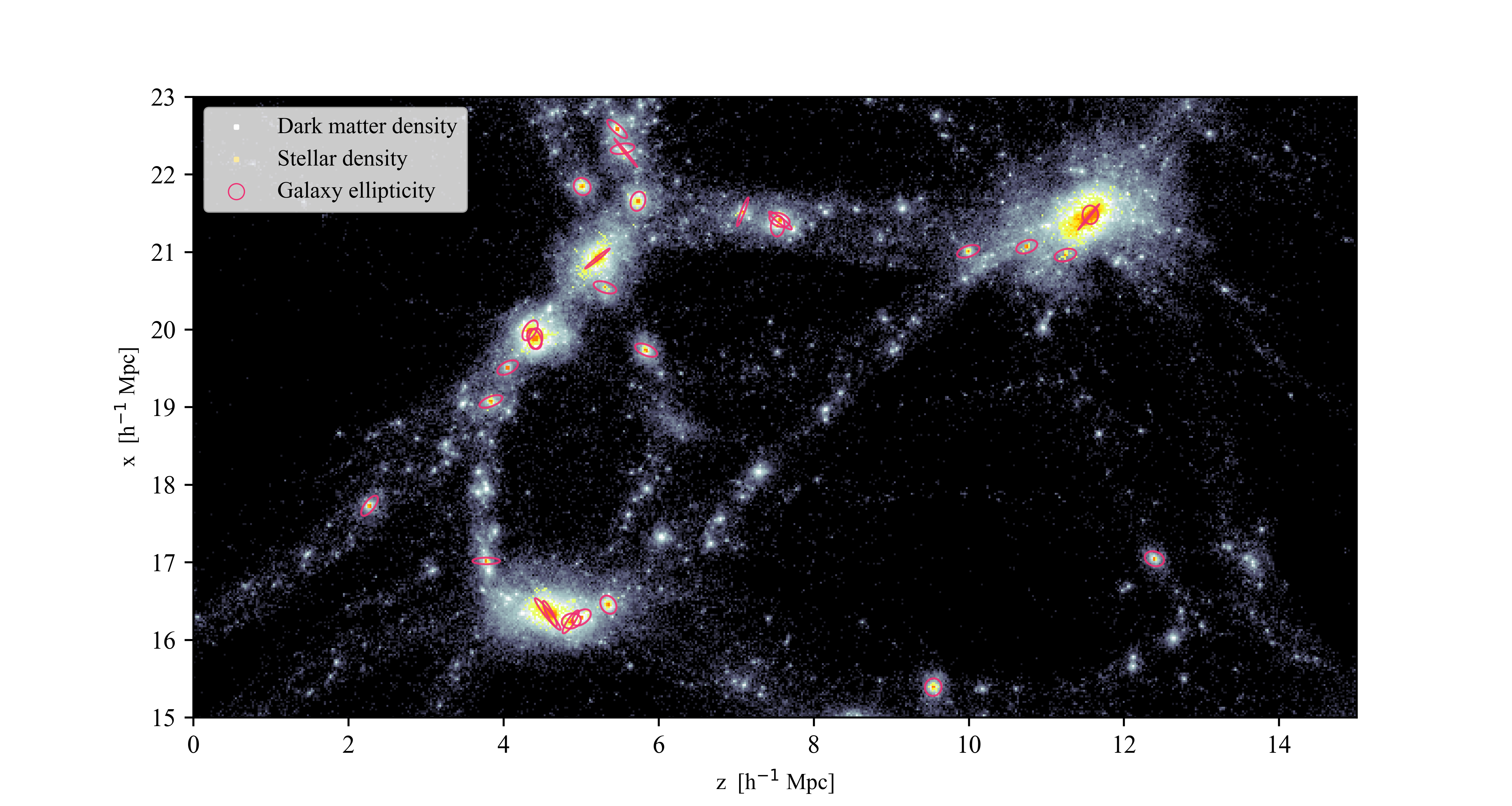}
    \caption{{A $1.67\,h^{-1}\,\text{Mpc}$ thick slice of the CAMELS LH-310 simulation viewed along the $y$-axis. The background shows the log-normal dark matter density with stellar density overlaid. Galaxy shapes are represented as ellipses with eccentricity and orientation encoding the magnitude and angle of the measured ellipticity $\varepsilon$. On average, galaxy shapes tend to align preferentially with the principal axes of the dark matter distribution, though intrinsic alignments in observational data are only detected statistically and not at the level of individual galaxies.}}
    \label{fig:ell_visual}
\end{figure*}
\section{Intrinsic alignment theory}\label{sec:Intrinsic alignment theory}

The aim of this work is not to explicitly model any particular theoretical prediction for the alignment signal. Instead, we provide a model-agnostic detection of the alignment signal and demonstrate its dependence on the astrophysical and cosmological parameters via simulations. Nevertheless, our choice of measured statistics are motivated by direct detection measurements from observations and, therefore, inspired by the theoretical models typically assumed for intrinsic alignments. 

We follow the approach taken by direct detection measurements from galaxy {survey data~\citep{Mandelbaum2006, 2011A&A...527A..26J, 2019Johnston, 2021A&A...654A..76F, 2023MNRAS.524.2195S, siegel2025} by} measuring alignments via the projected cross-correlation function $w_{m+}$ as a function of the comoving transverse separation $r_\perp$. Note that we use the subscript $m$ notation to refer to our correlation with respect to matter particles in the simulation, rather than galaxies $g$. This is given as a radial projection of the anisotropic two-point correlation function:
\begin{equation}\label{eq:wm+}
    w_{m+}(r_\perp)=\int^{\Pi_{\text{max}}}_{-\Pi_{\text{max}}}\xi_{m+}(r_\perp,\Pi)\,d\Pi \ \ ,
\end{equation}
\noindent for which we define our estimator in Section~\ref{sec:correlation functions}. $\Pi$ is the line of sight separation.

To aid interpretation of the results, recall that the expected signal $\bar{w}_{m+}$ is related to the matter-alignment power spectrum $P_{\delta I}(k, z)$ at a particular redshift $z$:
\begin{equation}\label{eq:wm+_pk}
\begin{split}
\bar{w}_{m+}(r_\perp, z) = \int_0^\infty \frac{\mathrm{d}k_z  }{\pi^2} \int_0^\infty \mathrm{d}k_\perp \, \frac{k_\perp}{k_z} \, P_{\delta I}(k, z) \, \times \\ \sin(k_z \Pi_{\mathrm{max}}) \, J_2(k_\perp r_\perp),
\end{split}
\end{equation}
for perpendicular $k_\perp$ and $k_z$ radial Fourier components. This assumes no redshift-space distortions~\citep{1987MNRAS.227....1K} or shape projection effects~\citep[see][]{2015MNRAS.450.2195S}.

In this work, we do not explicitly model $P_{\delta I}$ – the aim is provide a model-agnostic detection of the signal and its dependence on the cosmological and astrophysical parameters. However, in the literature there are a number of popular proposed prescriptions for $P_{\delta I}$, including the linear and non-linear alignment (NLA) models \citep{catelan2001, Hitara2004, Bridle_2007}, the tidal alignment and tidal torquing (TATT) model \citep{2019PhRvD.100j3506B}, and halo-based models \citep{2021MNRAS.501.2983F}.

Though we do not use these models, the form of the the simplest model motivates our choice to explore the ratio $\frac{w_{m+}}{w_{mm}}$. In the NLA model, $P_{\delta I}$ at redshift zero is given by
\begin{equation}\label{eq:power spec}
    P_{\delta I}(k,z=0) = -A_{IA} \,c_1\,\rho_{\rm crit}\Omega_m P_\delta(k,z=0),
\end{equation}
where $c_1$ is a fixed constant (chosen to rescale $A_{IA}$), $\rho_{\rm crit}$ is the critical density, and $\Omega_m$ is the matter density parameter. Recall that the expected matter autocorrelation is given by:
\begin{equation}\label{eq:wmm_pk}
\begin{split}
\bar{w}_{mm}(r_\perp, z) = \int_0^\infty \frac{\mathrm{d}k_z  }{\pi^2} \int_0^\infty \mathrm{d}k_\perp \, \frac{k_\perp}{k_z} \, P_{\delta}(k, z) \, \times \\ \sin(k_z \Pi_{\mathrm{max}}) \, J_0(k_\perp r_\perp),
\end{split}
\end{equation}
with matter power spectrum $P_\delta$. In the linear regime (at large scales) the ratio $\frac{w_{m+}}{w_{mm}} \appropto \Omega_m A_{IA}$ under the NLA model. That is, under assumptions of linearity, the ratio {will be less dependent than $w_{m+}$ on the cosmological parameters via the matter power spectrum (there is a residual effect due to the different $J_0$ and $J_2$ Bessel functions – we validate this in Appendix~\ref{sec:ratio_dependence})}.

In Section~\ref{sec:method}, we present the simulated alignment observables: the simulation suite, the shape measurement method, as the estimation of the correlation functions . In Section~\ref{sec:results}, we present: the detection of the alignment signal in CAMELS, the dependence on simulation parameters, the dependence on star formation rate, and the impact via the intrinsic ellipticity (via orientation-only correlation functions). We conclude in Section~\ref{sec:conclusion}.

\section{Simulated galaxy alignment observables}\label{sec:method}

\subsection{The CAMELS simulations}
The CAMELS simulations\footnote{https://www.camel-simulations.org/} are sets of hydrodynamical and N-body cosmological simulations that sample a high-dimensional parameter volume in order to explore connections between cosmology and astrophysical processes. We measure galaxy shapes in the \textit{Latin-Hypercube }(LH) set of simulations that use the hydrodynamic IllustrisTNG code. This set contains 1000 simulations, each having different values of selected cosmological and astrophysical parameters as detailed in Table~\ref{tab:camels_params}. The IllustrisTNG cosmic structure and galaxy formation model uses the \textsc{AREPO} gravity code \citep{springel2010, Vogelsberger2012} and incorporates various astrophysical processes such as baryonic effects detailed in \cite{Vogelsberger2013, Pillepich2017}.

All hydrodynamical simulations in CAMELS contain $256^3$ dark matter particles and $256^3$ fluid elements, and follow time evolution from redshift $z = 127$ down to $z = 0$ in a periodic box of volume $(25\,h^{-1}\,\text{Mpc})^3$. The simulations also assume a spatially flat cold dark matter (CDM) cosmological model with constant parameters: baryon density $\Omega_b=0.049$, Hubble parameter $h=0.6711$, dark matter equation of state $w=-1$, and sum of neutrino masses $\sum m_\nu=0$. While the zero neutrino mass is unphysical, we expect this common approximation to have negligible impact on our results. Further details on the effects of these parameters on the simulations can be found in \citet{Villaescusa-Navarro_2023}.

\begin{table}
    \centering
    \resizebox{\columnwidth}{!}{ 
    \begin{tabular}{l c c l}
        \hline
        \textbf{Parameter} & \textbf{Sampling} & \textbf{Range} & \textbf{Description} \\
        \hline
        $\Omega_m$  & Uniform  & $[0.1, 0.5]$ & Total matter density parameter \\
        $\sigma_8$  & Uniform  & $[0.6, 1.0]$ & Amplitude of matter fluctuations \\
        $A_{\text{SN1}}$  & Log-Uniform  & $[0.25, 4.0]$ & SN feedback energy injection \\
        $A_{\text{AGN1}}$ & Log-Uniform   & $[0.25, 4.0]$ & AGN feedback energy injection \\
        $A_{\text{SN2}}$  & Log-Uniform   & $[0.5, 2.0]$ & SN feedback wind velocity \\
        $A_{\text{AGN2}}$ & Log-Uniform   & $[0.5, 2.0]$ & AGN feedback wind velocity \\
        \hline
    \end{tabular}
    }
    \caption{CAMELS \textit{Latin-Hypercube} sampled parameters in IllustrisTNG suite. SN = Supernova, AGN = Active Galactic Nuclei.}
    \label{tab:camels_params}
\end{table}

\begin{figure}
	\includegraphics[width=\columnwidth]{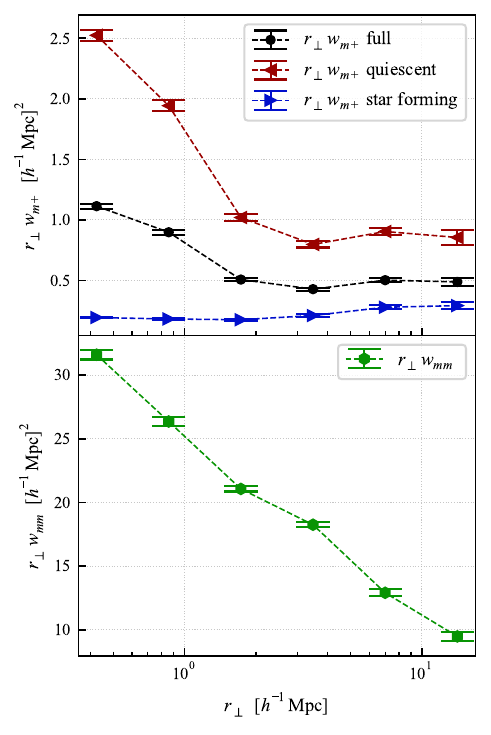}
    \caption{\textit{Top}: tangential-component correlation function averaged over all simulated catalogues $w_{m+}(r_\perp)$ for three galaxy samples: all galaxies, quiescent galaxies, and star forming galaxies. \textit{Bottom}: dark matter auto-correlation function $w_{mm}(r_{\perp})$ averaged over all simulated catalogues. $r_\perp$ values are plotted at the logarithmic centre of each of the six bins and a linear interpolation is added to guide the eye. All error bars are the standard error from $N_{\text{cat}}=3000$ catalogues.
    \label{fig:mean_corr1}}
\end{figure}

\subsection{Simulated shape catalogue}

For each simulation, stellar and dark matter particle positions at redshift $z=0$ were recorded together with the subhalo catalogue identified by the \textsc{SUBFIND} algorithm \citep{Dolag_Springer, springel_2001}.
Galaxies were selected from subhaloes based on the following criterion:
\begin{itemize}
    \renewcommand\labelitemi{}
    \item \hspace{-20pt} \textbf{Galaxy Criterion:} Let $M_{2r}$ be the stellar mass within twice the stellar half-mass radius and let $N_{\text{s}}$ be the number of stellar particles. Subhaloes with $M_{2r}>10^{8}\,M_\odot$ and $N_{\text{s}}>10$ are selected as our simulated galaxy sample.
\end{itemize}
The measured ellipticity of galaxies is two-dimensional and defined in projection. Therefore, three shape catalogues can be obtained from a single simulation box by measuring galaxy shapes along each of the three Cartesian projections, yielding a total of $N_{\text{cat}}=3000$ catalogues. For each projection, the galaxy shape $\varepsilon$ was calculated from the tensor of second moments $Q_{ij}$:
\begin{equation}\label{eq:epsillon}
    \varepsilon \equiv \frac{Q_{11}-Q_{22}+2iQ_{12}}{Q_{11}+Q_{22}+2(Q_{11}Q_{22}-Q_{12}Q_{21})^{1/2}}.
\end{equation}
Here $Q_{pq}$ is defined similarly to Eq.~{4.2} from \citet{BARTELMANN2001291}, but using stellar density instead of brightness (flux):
\begin{equation}
    Q_{pq} = \frac{\sum_{n=1}^{N_s}m_n \Delta_{pn}\Delta_{nq}}{\sum_{n=1}^{N_s}m_n}, \qquad p,q\in \{1,2\},
\end{equation}
$\Delta_{nj} = (x_{nj}-\bar x_{j})$ is the displacement vector to stellar particle $n$ from the galaxy's SUBFIND centre $\bar x_j$, $m_n$ is the mass of stellar particle $n$ and $N_s$ is the number of stellar particles that make up the galaxy. {We measure ellipticities in periodic space to avoid stellar particles being erroneously wrapped to the opposite side of the simulation box relative to $\bar{x}_j$.}

{Projected positions and ellipticities of galaxies are recorded from the $+z$, $+y$, and $+z$ coordinates allowing us to construct $N_\text{cat}=3000$ galaxy catalogues from 1000 simulations}. For runtime considerations, positions from a 1\% random sample of dark matter particles were taken to make up the corresponding dark matter catalogues.

{Figure~\ref{fig:ell_visual} shows an example slice of the dark matter density and the stellar density with measured ellipticities overlaid. For visualization we plot ellipses with semi-major axis $a=1+|\varepsilon|/2$, semi-minor axis $b=1-|\varepsilon|/2$, and orientation $\arg(\varepsilon)/2$.}

Dark matter and galaxy positions from our catalogues were transformed from Cartesian coordinates to right ascension and declination (RA, DEC) coordinates by positioning the centre of the simulation box at $10,000\,h^{-1}$ Mpc co-moving distance from the simulated observer. This setup enables the use of the two-point correlation function measurement pipeline developed in \citet{2019Johnston}, which employs \textsc{TreeCorr} \citep{treecorr} to compute the projected correlation functions $w_{m+}(r_\perp)$, $w_{m\times}(r_\perp)$, and $w_{mm}(r_\perp)$. {The coordinate transformation was performed to implement the pipeline and the positioning value of $10,000\,h^{-1}$ Mpc was chosen to minimise the error between separations in RA DEC coordinates and the Cartesian projected separations\footnote{The coordinate transform generates a $0.3\%$ maximum distance error relative to box size.}.}

\subsection{Correlation functions}\label{sec:correlation functions}
Recall that $r_\perp$ is the co-moving transverse separation and $\Pi$ is the line of sight separation. Our estimator for the matter density correlation is given by
\begin{equation}\label{eq:ddestimator}
    \xi_{mm}(r_{\perp},\Pi)=\frac{DD-RR}{RR}
\end{equation}
\citep{Peebles1974}, where $DD$ is the number of dark matter–dark matter pairs in bins of transverse and line-of-sight separation $(r_\perp,\Pi)$, and $RR$ is the number of such pairs in a corresponding randomly generated catalogue. Random catalogues are constructed with the same number of simulated dark matter particles as the original data. Given this high number of particles, the Poisson noise is much smaller than the signal. We do not use the Landy-Szalay estimator as our catalogues are spatially uniform and {the depth of variation becomes very small for the simulation box placed at high redshift.}

We compute the correlation between dark matter and intrinsic shear using an estimator similar to that in \cite{Mandelbaum2006}: 
\begin{equation}\label{eq:d+estimator}
    \xi_{m+}(r_\perp,\Pi)=\frac{S_+D-S_+R}{R_SR},
\end{equation}
where $S_+D$ is the sum over all dark matter–galaxy pairs of the tangential component of galaxy ellipticity, at ($r_\perp$, $\Pi$):
\begin{equation}\label{eq:g_dm pairs}
    S_+D=\sum_m\sum_g\varepsilon_+(g|m).
\end{equation}
Here, $\varepsilon_+(g|m)$ is the tangential component of galaxy $g$ relative to the direction of dark matter particle $m$. A value $0<\varepsilon_+(g|m)\le1$ indicates radial alignment or elongation towards simulated dark matter and $-1\le\varepsilon_+(g|m)<0$ indicates tangential
alignment. The term $S_+R$ is computed analogously using random positions for the dark matter field. The normalization factor $R_SR$ counts the number of dark matter-galaxy pairs in the random catalogue. 

To validate our measurement, we also compute the cross-component correlation function, which serves as a null test:
\begin{equation}\label{eq:dxestimator}
\xi_{m\times}(r_\perp,\Pi) = \frac{S_\times D - S_\times R}{R_SR}.
\end{equation}
This is calculated in the same way as $\xi_{m+}$ but using the cross ellipticity component $\varepsilon_\times(g|m)$ in place of $\varepsilon_+$.

We considered pairwise signals for the 2D correlation functions (Eqs.~{\ref{eq:ddestimator}},~{\ref{eq:d+estimator}} and~\ref{eq:dxestimator}) in co-moving range $|\Pi| \le 12 \,h^{-1}$ Mpc discretized into 24 bins of $\Delta\Pi=1 \,h^{-1}$ Mpc. {We note that $\Pi_{\text{max}}$ is much smaller than typical $\Pi_\text{max}$ values used in surveys due to our limited simulation volume \citep{Mandelbaum2006, 2011A&A...527A..26J, 2019Johnston}.} Transverse separations $r_\perp$ were considered in six logarithmically spaced bins in the range $0.3 \le r_\perp \le 20\,h^{-1}$ Mpc. We sum over $\Delta\Pi$ intervals to approximate the integral in Eq. \ref{eq:wm+} and obtain projected correlation functions $w_{m+}(r_\perp)$, $w_{m\times}(r_\perp)$ and $w_{mm}(r_\perp)$.

With our estimate of $w_{mm}$, the small volume of the CAMELS simulations induces an additive bias through the \textit{integral constraint} \citep{Peebles:1980yev} – this is a common observation effect in small galaxy surveys~\citep[e.g.][]{1993MNRAS.263..360R, 1999MNRAS.306..988M}. In our case, if uncorrected, we measure negative values for  $w_{mm}$ at high $r_\perp$. Our correction only affects results that use the ratio $\frac{w_{m+}}{w_{mm}}$, and we find that the conclusions are not affected by the exact value. 

{
We correct our estimates for the integral constraint bias of $w_{mm}$ with an additive constant $C=1$. The shape-density correlation $w_{m+}$ is unaffected. As the value of $C$ is formally signal dependent, in Appendix~\ref{sec:integral_constraint_correction} we fit a model per simulation to estimate a per simulation value of $C$, which we use to define our approximate value $|C| \approx 1$. The final conclusions are not affected by the value of $C$.}

We split the galaxy catalogues at a specific star formation rate (sSFR) of $10^{-10.5}$ /yr \citep{L_Bisigello_Euclid_split} and make three samples: full, star forming, and quiescent. Note that in IllustrisTNG, the galactic SFR is defined as the sum of star formation rates of all gas cells existing in the subhalo.

The top panel in Figure~\ref{fig:mean_corr1} shows the mean tangential matter-shear correlation functions $w_{m+}(r_{\perp})$ over all galaxy catalogues for our three samples. The bottom panel similarly shows the mean dark matter density auto-correlation $w_{mm}(r_{\perp})$ calculated from dark matter catalogues while Figure~\ref{fig:mean_corr_wdx} displays the cross-correlation function $w_{m\times}(r_{\perp})$. {Error bars for all mean correlation functions are the standard error for $N_\text{cat}=3000$ catalogues.}

\subsection{{N}ormalizing the magnitudes of ellipticities}\label{sec:ell norm}
Galaxy shapes encode both the orientation and the magnitude of ellipticity. {To isolate the orientations of galaxies we recalculate the correlation functions from Section~{\ref{sec:correlation functions}} with ellipticity magnitudes set to a constant $\langle |\varepsilon|\rangle$:}

\begin{equation}
    \langle|\varepsilon|\rangle=\frac{\sum^{N_{\text{cat}}}\limits_{i=1}\sum\limits^{N_i}_{g=1}|\varepsilon_{ig}|}{\sum\limits^{N_{\text{cat}}}_{i=1}N_{i}}.
\end{equation}
Here, $N_i$ is the number of galaxies in catalogue $i$ and $N_\text{cat}$ is the number of catalogues. 

\section{Alignment detection and dependence}\label{sec:results}
\subsection{Alignment signal from catalogues}
\begin{figure}
	\includegraphics[width=\columnwidth]{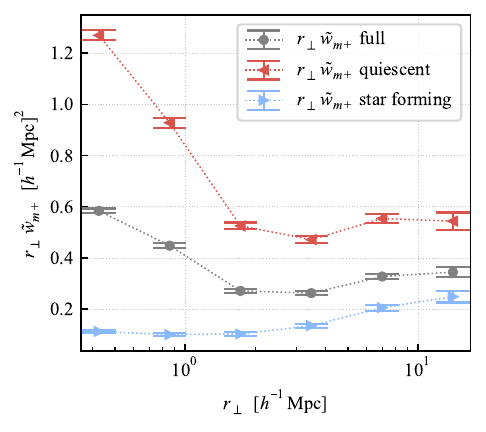}
    \caption{Orientation-only correlation functions averaged over all simulated galaxy catalogues $\tilde w_{m+}(r_\perp)$, who's catalogues have the magnitudes of galaxy ellipticities set to constant $\langle |\varepsilon|\rangle$. Full, quiescent and star forming galaxy samples are displayed. Correlation functions here take the same shape as Figure~\ref{fig:mean_corr1} indicating that the magnitudes of the ellipticities predominantly impact alignment amplitude. Linear interpolation is added to guide the eye{.}}
    \label{fig:mean_corr2}
\end{figure}
{Figure~\ref{fig:mean_corr1} demonstrates the detection of intrinsic alignment of galaxies in the CAMELS simulations. The result is obtained by taking the mean projected correlation over all $N_{\text{cat}}$ catalogues.}

We also demonstrate that the alignments of quiescent galaxies are significantly stronger than those of high star forming galaxies in our CAMELS catalogues. Again in Figure~\ref{fig:mean_corr1}, $w_{m+}(r_\perp)$ for the sample of galaxies below sSFR $10^{-10.5}$ yr$^{-1}$ is consistently higher than that of the high star forming sample. This agrees with previous studies showing that alignments behave differently in early and late-type galaxies {\citep{hirata2007, mandelbaum2011, samuroff_sim}. We note that the star-forming population may be contaminated by late-type galaxies as some models of IA \citep{chisari_15, 2019PhRvD.100j3506B} predict a negative (tangential) alignment for high star-forming galaxies.}

Figure~\ref{fig:mean_corr2} presents mean measured orientation-only correlation functions, denoted $\tilde{w}_{m+}(r_\perp)$ calculated from orientation-only catalogues outlined in Section~\ref{sec:ell norm}. As before, mean functions from the full, quiescent and star forming samples are taken. {We find that normalizing ellipticity magnitudes affects alignments uniformly across $0.3\le r_\perp \le 20 \, h^{-1}$ Mpc as the functions $w_{m+}(r_\perp)$ and $\tilde w_{m+}(r_\perp)$ (in Figures~\ref{fig:mean_corr1} and \ref{fig:mean_corr2} respectively) have similar scale dependence. This is a feature found in intrinsic alignment models \citep{Schneider_2010, Blazek_2011} where the scale dependence is set by the tidal field. This result has also been observed previous works on surveys such as \citet{2024MNRAS.534..3540}.}

\subsection{Validation}\label{sec:validation}
\begin{figure}
	\includegraphics[width=\columnwidth]{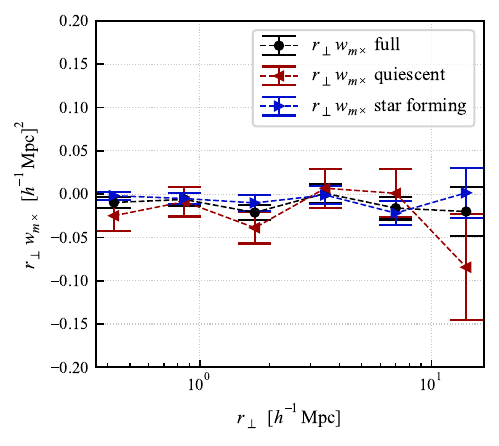}
    \caption{Cross-component correlation function averaged over all catalogues $w_{m\times}(r_\perp)$ for full, quiescent and star forming galaxy samples. This null-test measurement is consistent with zero, as expected. Linear interpolation is added to guide the eye.}
    \label{fig:mean_corr_wdx}
\end{figure}

In line with previous studies of intrinsic alignments \citep{Mandelbaum2006, hirata2007, 2011A&A...527A..26J, 2019Johnston}, we validate that the detected alignment signal arises from astrophysical effects rather than systematic errors introduced by the measurement pipeline or simulation setup.

We compute the mean position shape cross-component correlation function $w_{m\times}(r_\perp)$, alongside $w_{m+}(r_\perp)$, as a null test. This statistic corresponds to rotating all catalogue galaxies' shapes by 45 degrees and recomputing $w_{m+}(r_\perp)$. Since $w_{m\times}(r_\perp)$ changes sign under parity (rotating galaxies clockwise or anti-clockwise), it must be zero to preserve the parity invariance of galaxy formation. The measured $w_{m\times}(r_\perp)$ for both full and split galaxy samples is shown in Figure~\ref{fig:mean_corr_wdx}, with error bars denoting the standard error on the mean. All measurements are consistent with zero, indicating no significant systematic bias in our $w_{m+}$ signal. We note the presence of B-modes may provide a small, non-zero $w_{m\times}(r_\perp)$ signals. It has been proposed that intrinsic alignments may be sensitive to such B-mode contributions \citep{Georgiou_2024}, which we do not detect in these simulations.

As an additional null test, we randomly rotate galaxy shapes in each catalogue and confirm that both $w_{m+}$ and $w_{m\times}$ vanish, as expected for uncorrelated orientations. These tests support the robustness of our alignment measurements.

\subsection{Dependence on simulation parameters}\label{sec:param dep}
\begin{figure*}
    \centering
    \includegraphics[width=0.9\linewidth]{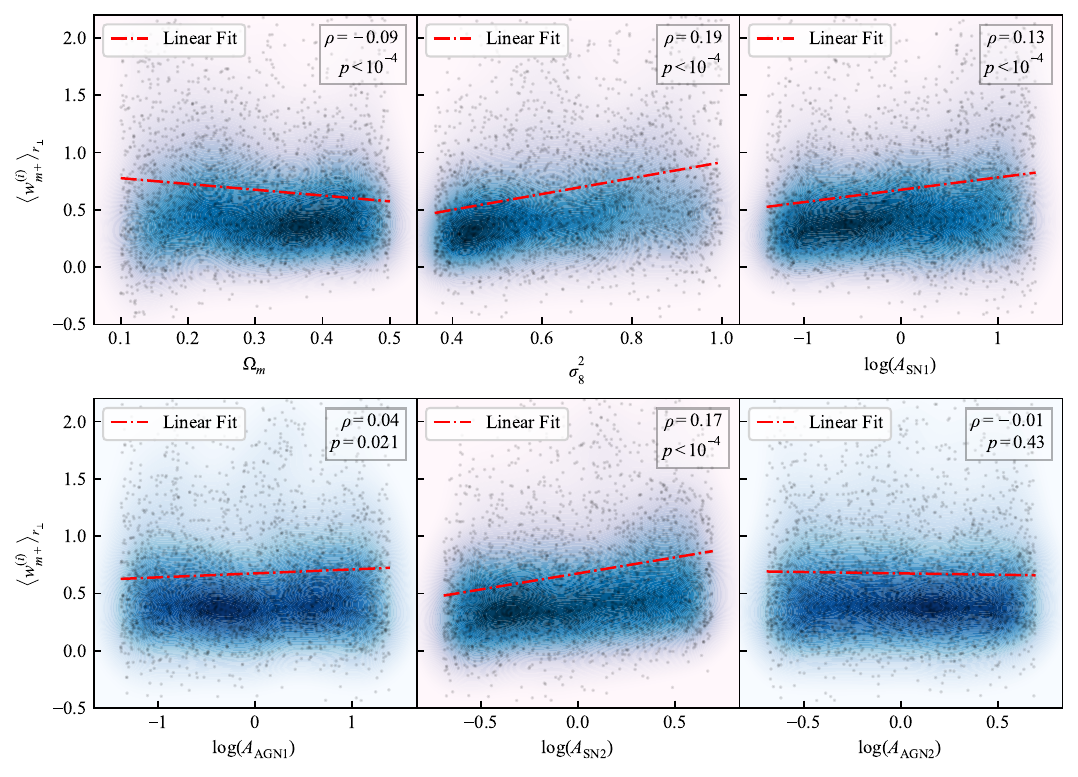}
    \caption{Correlation functions weighted and averaged over transverse distances $0.3\le r_\perp\le20$, $\langle w^{(i)}_{m+} \rangle_{r_\perp}$, versus each of six simulation parameters for all simulation catalogues $i$. A density histogram overlays the scatter plot and a linear fit with Pearson value $\rho$ measures linear correlation. The $p$-value $p<10^{-4}$ suggests a statistically significant correlation between gravitational alignments and simulation parameter. Note: the $\langle w^{(i)}_{m+} \rangle_{r_\perp}$ distribution over $i$ has a long tail to high values, so the red linear fit positioned at the mean is not aligned with the mode.}
    \label{fig:dep_wdp}
\end{figure*}
We present the dependence of the astrophysical and cosmological parameters on measured galaxy alignments. Alignment signals are analysed across our $N_{\text{cat}}$ catalogues—each generated with different set of simulation parameters listed in Table~\ref{tab:camels_params}.

Let $w_{m+}^{(i)}(r_{\perp,j})$ denote the measured shape-density correlation for the $i$-th catalogue in the $j$-th $r_\perp$ bin, with $ j = 1, \dots, 6 $.
We define the alignment amplitude of catalogue $i$ as the weighted average of $w_{m+}^{(i)}(r_{\perp,j})$ over $r_\perp$ bins:
\begin{equation}\label{eq:ali amp}
    \big\langle w_{m+}^{(i)} \big\rangle_{r_\perp} = \frac{\sum\limits_{j=1}^{6} w_j \, w_{m+}^{(i)}(r_{\perp,j})}{\sum\limits_{j=1}^{6}w_j}
\end{equation}
where the weights $w_j$ are given by the inverse of the standard error on the mean {for the $j$'th bin over all catalogues}. This weighting scheme down-weights bins with higher variance across catalogues, reducing the influence of noisy measurements in the final alignment amplitude.

For each catalogue $i$, we compute the alignment amplitude $\langle w_{m+}^{(i)} \rangle_{r_\perp}$ and display these values in the six subplots in Figure~\ref{fig:dep_wdp}, against simulation parameters. A logarithmic scaling is applied to $A_{\text{SN1}}$, $A_{\text{SN2}}$, $A_{\text{AGN1}}$, and $A_{\text{AGN2}}$ to ensure these parameters are uniformly distributed. We measure correlation with $\sigma_8^2$ because $w_{m+}$ is approximately proportional to $\sigma_8^2$, not $\sigma_8$, under the linear approximation described in Section~\ref{sec:Intrinsic alignment theory} \citep{Singh_2016}. Linear fits are overlaid on scatter and density plots from the sample means to guide the eye.

We employ Pearson coefficients and $p$-values to measure linear correlations between alignment amplitude and simulation parameters. {The statistical significance of each correlation is measured by calculating} a two-tailed p-value associated with the null hypothesis $H_0$: that there is no linear correlation between $\langle w_{m+}^{(i)} \rangle_{r_\perp}$ and {the simulation parameter} (i.e., $\rho=0$). The $p$-value is computed under the assumption that the test statistic {$t$}
follows a Student's $t$-distribution with $N_{\text{cat}}-2$ degrees of freedom.

We adopt a significance threshold of $p < 10^{-4}$ to reject the null hypothesis. Correlations with $p$-values below this threshold are considered statistically significant, indicating that $w_{m+}$ is dependent on the simulation parameter under consideration. Subplots in Figure~\ref{fig:dep_wdp} are annotated with Pearson coefficients and $p$-values for each parameter–alignment amplitude pair.

We measure statistically significant correlations for $\log A_{\text{SN1}}$ and $\log A_{\text{SN2}}$, implying that supernova feedback influences the alignment of galaxies at the $r_\perp$ scales considered.

We further observe significant correlations with cosmological parameters $\Omega_m$ and $\sigma_8^2$. Alignment dependence on the large-scale structure and these two parameters is predicted in prescriptions of alignment theory discussed in Section~\ref{sec:Intrinsic alignment theory}.

No significant correlation is detected between alignment amplitude $\langle w_{m+}^{(i)} \rangle_{r_\perp}$ and either $\log A_{\text{AGN1}}$ or $\log A_{\text{AGN2}}$. {A lack of AGN dependence is likely due to the $(25\,h^{-1}\,\text{Mpc})^3$ simulation volume that limits the full range of large-scale modes that influence intrinsic alignments in cosmological surveys \citep{Pandey2023}. Other simulation studies \citep{tenneti2015, Soussana2020} show that AGN feedback has negligible impact on matter-shape correlations but may impact galaxy-shape correlations especially when considering high-mass halos.}

\subsection{Parameter dependence of ratio $\frac{w_{m+}}{w_{mm}}$}\label{sec:wm+wmm}
\begin{figure}
	\includegraphics[width=\columnwidth]{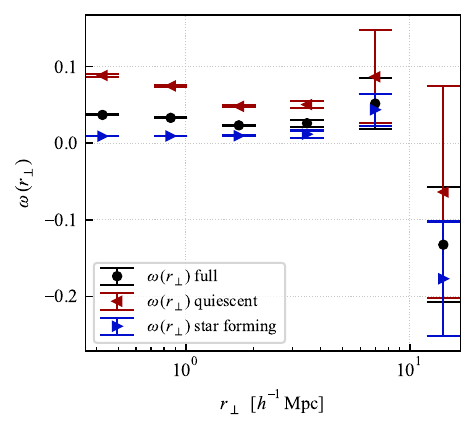}
    \caption{{Ratio of the density–shape to density–density correlation functions, averaged across all catalogues: $\omega(r_{\perp})=\langle\omega^{(i)}(r_{\perp})\rangle$. Mean ratios are shown for the full, quiescent, and star-forming galaxy samples.}}
    \label{fig:wdp/wdd}
\end{figure}
In this section we explore the dependence of ratio $\frac{w_{m+}}{w_{mm}}$ on simulation parameters using the same methods as Section~\ref{sec:param dep}.

{Let $\omega^{(i)}(r_\perp)$ be the ratio of the $w_{m+}^{(i)}(r_{\perp})$ and $w_{mm}^{(i)}(r_{\perp})$ correlation functions for the $i$-th catalogue:
\begin{equation}\label{eq:ratio}
    \omega^{(i)}(r_\perp) = \frac{w_{m+}^{(i)}(r_{\perp})}{w_{mm}^{(i)}(r_{\perp})}
\end{equation}
Figure~\ref{fig:wdp/wdd} displays the mean ratios ${\omega}(r_\perp)$ for our samples averaged over all $N_{\text{cat}}$ catalogues with standard error on the mean errors.}

Similarly to Section~\ref{sec:param dep}, we detect statistically significant correlations between {the alignment amplitude of ratio $\omega^{(i)}(r_\perp)$ (for the $i$-th catalogue) and simulation parameters. As in Eq.~\ref{eq:ali amp}, we take the weighted average over $r_\perp$ with weights equal to the inverse of the standard error on the mean.}

\begin{figure*}
    \centering
    \includegraphics[width=\linewidth]{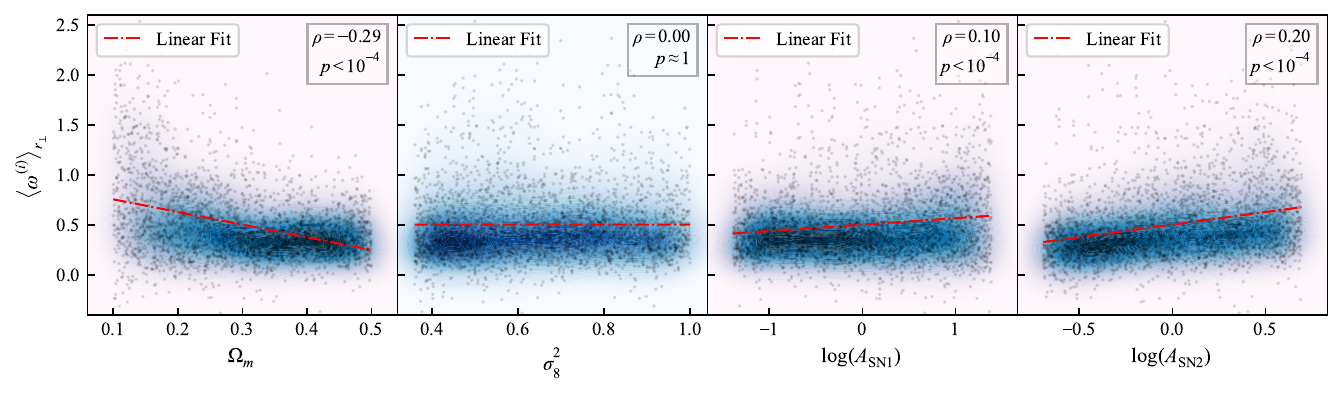}
    \caption{The same as Figure~\ref{fig:dep_wdp} but testing the weighted ratio {$\langle\omega^{(i)}\rangle_{r_\perp}$}, with catalogue index $i$, for parameters $\Omega_m$, $\sigma_8^2$, $\log A_{\text{SN}1}$, $\log A_{\text{SN}2}$.}
    \label{fig:dep_wdp/wdd}
\end{figure*}

Figure~\ref{fig:dep_wdp/wdd} displays the weighted average ratio {$\langle\omega^{(i)}\rangle_{r_\perp}$} against each simulation parameter, following the same format as Figure~\ref{fig:dep_wdp}. Linear regression lines are overlaid to illustrate trends. For each parameter–{$\langle\omega^{(i)}\rangle_{r_\perp}$} pair, the Pearson correlation coefficient and the corresponding $p$-value is calculated to assess the statistical significance of the relationship against the null hypothesis $H_0$ of no correlation.

We detect no significant correlation between measured {$\langle\omega^{(i)}\rangle_{r_\perp}$} and $\sigma_8^2$ indicating that dividing by $w_{mm}$ removes the (linear) amplitude density fluctuation dependence on $w_{m+}$ in $\frac{w_{m+}}{w_{mm}}$. This result broadly follows predictions from NLA (see Appendix~\ref{sec:ratio_dependence})  {where most clustering dependence is removed by dividing by $w_{mm}$.}.

In contrast, {$\langle\omega^{(i)}\rangle_{r_\perp}$} retains statistically significant correlations with the matter density parameter $\Omega_m$. The dependence on $\Omega_m$ is not itself surprising. However, the negative correlation between {$\langle\omega^{(i)}\rangle_{r_\perp}$} and $\Omega_m$ not what one would expect from the NLA or TATT models (see Appendix~\ref{sec:ratio_dependence}). This is surprising, and merits further work. With the volume of the present generation of CAMELS simulations, we are not able to test this behaviour further into the linear regime.

We also find that {$\langle\omega^{(i)}\rangle_{r_\perp}$} retains statistically significant correlations with the supernova feedback parameters $\log A_{\text{SN}1}$ and $\log A_{\text{SN}2}$. No significant correlations are observed for $\log A_{\text{AGN1}}$ and $\log A_{\text{AGN2}}$ which are omitted from Figure~\ref{fig:dep_wdp/wdd}.

\subsection{Star formation rate and alignments}\label{sec:SFR and alignments}
 
\begin{table*}
\centering
\begin{tabular}{ccccccc c ccc}
\multicolumn{1}{c}{} 
& \multicolumn{3}{c}{$\big\langle w^{(i)}_{m+} \big\rangle_{r_{\perp}}$} 
& \multicolumn{3}{c}{$\big\langle \tilde w^{(i)}_{m+} \big\rangle_{r_{\perp}}$} 
& 
& \multicolumn{3}{c}{$\langle|\varepsilon|\rangle_i$} \\
\cmidrule(rl){2-4} \cmidrule(rl){5-7} \cmidrule(rl){9-11}
\textbf{Param} 
& {All} & \begin{tabular}{@{}c@{}}Star\\ form.\end{tabular} & {Quiescent} 
& {All} & \begin{tabular}{@{}c@{}}Star\\ form.\end{tabular} & {Quiescent}
&
& {All} & \begin{tabular}{@{}c@{}}Star\\ form.\end{tabular} & {Quiescent} \\
\cmidrule{1-7} \cmidrule{9-11}
$\Omega_m$        & \cellcolor{gbox}{-0.09} & \cellcolor{gbox}{-0.20} & \cellcolor{gbox}{-0.20} & \cellcolor{gbox}{0.08} & -0.06 & -0.07 & &  \cellcolor{gbox}{-0.75} & \cellcolor{gbox}{-0.75} & \cellcolor{gbox}{-0.66} \\
$\sigma_8^2$        & \cellcolor{gbox}{0.19}  & 0.02                    & \cellcolor{gbox}{0.10}  & \cellcolor{gbox}{0.22} & -0.03 & \cellcolor{gbox}{0.14} & & \cellcolor{gbox}{-0.14} & \cellcolor{gbox}{-0.17} & \cellcolor{gbox}{-0.13} \\
$\log A_{\text{SN}1}$ & \cellcolor{gbox}{0.13}  & \cellcolor{gbox}{0.30} & \cellcolor{gbox}{0.16}  & \cellcolor{gbox}{0.18} & \cellcolor{gbox}{0.30} & \cellcolor{gbox}{0.22} & & \cellcolor{gbox}{0.42} & \cellcolor{gbox}{0.48} & \cellcolor{gbox}{0.32} \\
$\log A_{\text{SN}2}$ & \cellcolor{gbox}{0.17}  & \cellcolor{gbox}{0.14} & -0.01                   & \cellcolor{gbox}{0.18} & \cellcolor{gbox}{0.16} & -0.06 & & 0.09 & -0.01 & 0.11 \\
\cmidrule[\heavyrulewidth]{1-7} \cmidrule[\heavyrulewidth]{9-11}
\end{tabular}
\caption{Pearson values for simulation parameters against: alignment amplitude for original catalogues $\langle w^{(i)}_{m+}\rangle_{r_\perp}$, alignment amplitude for ellipticity-normalized catalogues $\langle \tilde w^{(i)}_{m+}\rangle_{r_\perp}$, and average ellipticity magnitude of the sample $\langle|\varepsilon|\rangle_{\text{samp}}$, $i$ indexes the catalogue. Columns contain Pearson values for the full, star forming, and quiescent galaxy samples. Shaded cells indicate significant Pearson values with $p$-value $<10^{-4}$. Rows $\log A_{\text{AGN}1}$ and $\log A_{\text{AGN}2}$ showed no significant correlation (as expected given the small simulation size) are omitted from the table.}
\label{table:pearson_vals}
\end{table*}
Some parameter dependencies observed in the full galaxy sample may arise from underlying differences between star-forming and quiescent populations. When these subpopulations exhibit distinct behaviours, combining them can obscure or even reverse trends—a phenomenon known as Simpson’s paradox. We analyse how intrinsic alignments correlate with model parameters separately for each population. Pearson coefficients similar to those in Section~\ref{sec:param dep} were calculated for star-forming and quiescent samples. The left three sub-columns of Table~\ref{table:pearson_vals} presents the full set of Pearson coefficients with highlighted statistical significance.

Two notable observations {about our matter-shape correlations $w_{m+}$} emerge from these results: (1) correlation between alignment amplitude and $\sigma_8$ is insignificant for star-forming galaxy catalogues but strong for quiescent and full samples; (2) the correlation with $A_{\text{SN1}}$ is stronger in the star forming sample than in quiescent and full samples.
\paragraph*{$\sigma_8$:}
\begin{figure}
    \centering
    \includegraphics[width=\linewidth]{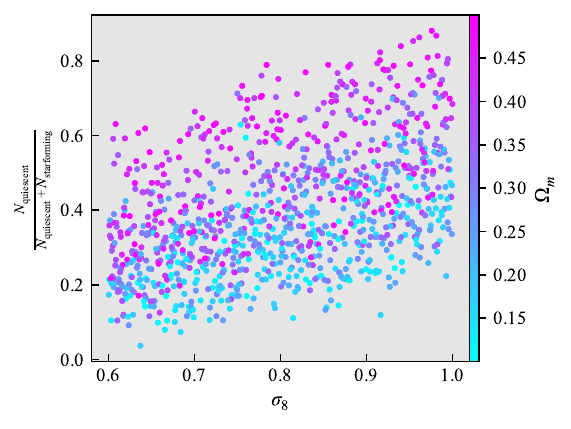}
    \caption{The fraction of quiescent galaxies $\frac{N_{\text{quiescent}}}{N_{\text{quiescent}}+N_{\text{star forming}}}$ against $\sigma_8$ parameter value for the catalogues, coloured by $\Omega_m$ parameter. Parameters $\sigma_8$ and $\Omega_m$ both positively correlate with a larger fraction of quiescent galaxies.}
    \label{fig:quiescent_prop}
\end{figure}
The $r_\perp$ averaged correlation function $\langle w^{(i)}_{m+}\rangle_{r_\perp}$ for the star forming sample shows no  significant correlation with the magnitude of matter density fluctuations $\sigma_8$, while $\langle w^{(i)}_{m+}\rangle_{r_\perp}$ for the quiescent sample displays a weaker correlation than in the full. 

Figure~\ref{fig:quiescent_prop} illustrates that increased $\Omega_m$ and $\sigma_8$ parameter values cause a larger proportion of quiescent galaxies in the catalogue. Since quiescent galaxies exhibit stronger alignments, by approximately an order of magnitude
\begin{equation}
\frac{E[\langle w_{m+}\rangle_{r_\perp}\text{ quiescent}]}{E[\langle w_{m+}\rangle_{r_\perp}\text{ star forming]}} \sim 10^1,
\label{eq:meanvals}
\end{equation}
the observed dependence on $\sigma_8$ in the full sample is amplified by the increasing proportion of quiescent galaxies at higher $\sigma_8$. On the contrary this effect produces a weaker negative dependence on $\Omega_m$.

Assuming a face-value NLA or TATT model, if the trend shown in Figure~\ref{fig:quiescent_prop} is accurate, then the result in Figure~\ref{fig:dep_wdp/wdd} is surprising, as it shows no $\sigma_8^2$ dependence in the ratio $\frac{w_{m+}}{w_{mm}}$. Specifically, Figure~\ref{fig:quiescent_prop} suggests that the full galaxy sample should exhibit not only the predicted linear $\sigma_8^2$ dependence (from the density field) but also an additional dependence due to the increasing fraction of quiescent galaxies at higher $\sigma_8$, which contribute more strongly to the alignment signal.

The inconsistency lies in the fact that the NLA and TATT models predict that the ratio $\frac{w_{m+}}{w_{mm}}$ should remove only the linear $\sigma_8^2$ scaling of $w_{m+}$ inherited from the density field, and would leave the sample-dependent $\sigma_8$ dependence (seen in Figure~\ref{fig:quiescent_prop}). However, the result in Figure~\ref{fig:dep_wdp/wdd} show that this ratio has cancelled out not only the $\sigma_8^2$ dependence, but also the additional $\sigma_8$ dependence arising from the changing presence of quiescent galaxies in the population.

\paragraph*{$A_{\text{SN1}}$:} 
\begin{figure}
	\includegraphics[width=0.95\columnwidth]{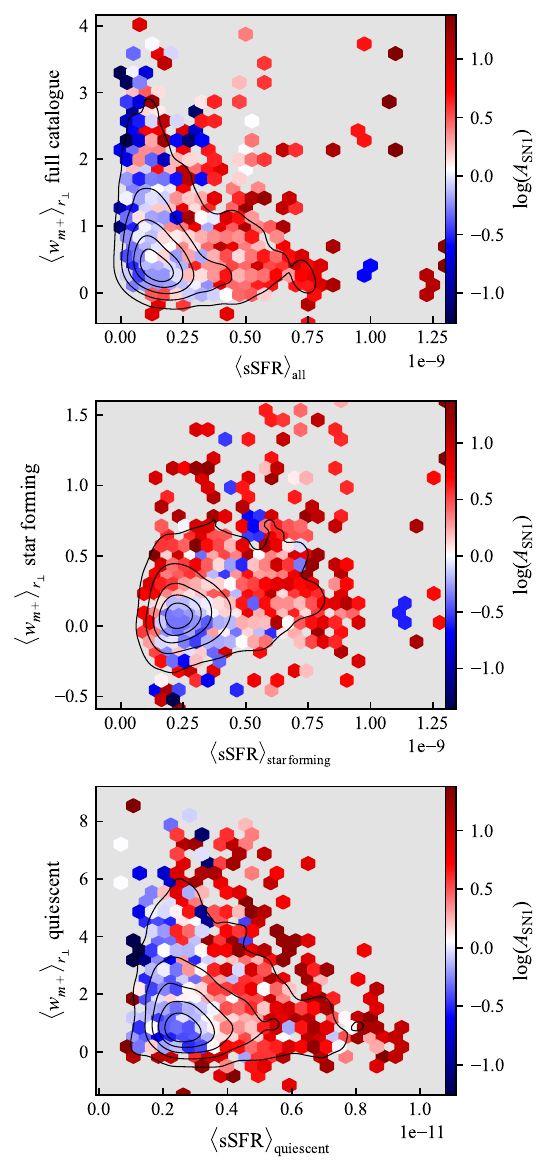}
    \caption{Alignment amplitude $\langle w_{m+}\rangle_{r_\perp}$ versus mean specific star formation rate $\langle \text{sSFR}\rangle_{\text{samp}}$ for our full (top), star forming (middle) and quiescent (bottom) galaxy samples. Each 2D bin is coloured by the average $\log (A_{\text{SN1}})$ of those sampled catalogues within it. A contour plot of point density shows the distribution of sampled catalogues.
    The shift in colour distribution between star-forming and quiescent samples indicates that $A_{\text{SN1}}$ influences intrinsic alignments differently across galaxy populations.}
    \label{fig:ssfr e ASN1}
\end{figure}
We observe a shift in galaxy dependence on $A_{\text{SN1}}$ between our two classes of galaxies. Figure~\ref{fig:ssfr e ASN1} displays alignment amplitude $\langle w^{(i)}_{m+}\rangle_{r_\perp}$ against the mean specific star formation rate of galaxies in each sample $\langle \text{sSFR} \rangle_{\text{samp}}$\footnote{The subscripts on angular brackets denote averaging over the respective galaxy sample: $\langle \cdot \rangle_{\text{full}}$, $\langle \cdot \rangle_{\text{star forming}}$, or $\langle \cdot \rangle_{\text{quiescent}}$.}. There is a clear change in the coloured $\log (A_{\text{SN1}})$ distribution between the quiescent (bottom) and star-forming (middle) plots indicating that the parameter's primary influence on galaxies shifts from star formation rate to alignment. Within star forming galaxies, catalogues with large $\log (A_{\text{SN1}})$ produced stronger galactic alignments to dark matter density, while the quiescent galaxies primarily exhibited increased specific star formation rates. As star forming galaxies contribute weak alignment relative to quiescent ones (Eq. \ref{eq:meanvals}), the top (full sample) plot follows a similar $\log (A_{\text{SN1}})$ dependency to the quiescent sample.

\subsection{Intrinsic ellipticity and alignment dependence}\label{sec: ell norm analyse}
We further investigate the parameter dependence of intrinsic alignments using the orientation-only galaxy catalogues defined in Sec.~\ref{sec:ell norm}. The corresponding correlation functions, $\tilde w^{(i)}_{m+}$, only register signal from galaxy orientations as all galaxy ellipticity magnitudes are normalized. As such, comparing the parameter dependencies of the original $w^{(i)}_{m+}$ and the orientation-only $\tilde w^{(i)}_{m+}$ correlation functions offers insight into the role played by ellipticity magnitude in shaping alignment signals and their dependencies.

Analogous to Eq.\ref{eq:ali amp}, we compute alignment amplitudes $\langle \tilde w^{(i)}_{m+} \rangle_{r_{\perp}}$ for the orientation-only correlation functions. We then calculate Pearson correlation coefficients between these amplitudes and the simulation parameters, with the values and their statistical significance presented in the centre column of Table~\ref{table:pearson_vals}. A comparison between the left and centre columns of Table~\ref{table:pearson_vals} highlights how the parameter dependence differs between the original and orientation-normalized galaxy samples.

Additionally, we compute the explicit correlations between the average absolute ellipticity $\langle |\varepsilon| \rangle_i$ of catalogue $i$ (of the original catalogues) and the simulation parameters, listed in the rightmost columns of Table~\ref{table:pearson_vals}. These correlations help identify parameter dependencies present in $\langle |\varepsilon| \rangle_i$, providing insight into the information lost when measuring alignments for orientation-only catalogues.

For parameter $\Omega_m$, Pearson values across all samples are more positive with $\langle \tilde w^{(i)}_{m+} \rangle_{r_{\perp}}$ than with $\langle w^{(i)}_{m+} \rangle_{r_{\perp}}$. We also identify a direct negative correlation between $\Omega_m$ and average ellipticity magnitude $\langle |\varepsilon| \rangle_i$. By removing the ellipticity magnitude, the negative $\tilde w^{(i)}_{m+}$ correlation with $\Omega_m$ is reduced. This suggests that the absolute ellipticity magnitude accounts for some of the $\Omega_m$ dependence in the correlation function $w^{(i)}_{m+}$. {Previous studies \citep{tenneti2015, Soussana2020} have shown that the scale dependence of alignment is unaffected by ellipticity magnitude. Our results are consistent with these findings: as shown in Figure~\ref{fig:mean_corr2}, normalizing the ellipticity magnitude simply rescales the amplitude of the correlation function $w_{m+}(r_\perp)$ by a constant factor.}

Interestingly, the supernova feedback dependence on the correlation function is retained with the orientation-only catalogues. Correlations between $\langle \tilde w^{(i)}_{m+} \rangle_{r_{\perp}}$ and $\log A_{\text{SN1}}$ are even stronger than those between $\langle w^{(i)}_{m+} \rangle_{r_{\perp}}$ and $\log A_{\text{SN1}}$ for full and quiescent samples. This is despite removing the positively correlated ellipticity magnitude. This demonstrates that supernova feedback ($A_{\text{SN1}}$) has a strong influence on the orientation of galaxies, not just on their ellipticity magnitudes. Increased $\log A_{\text{SN1}}$ must influence orientations such that they align more strongly to dark matter density. The reason for the strengthened correlation, and the exact mechanism linking $A_{\text{SN1}}$ to intrinsic alignment, remain open questions for future investigation.

\section{Conclusions} \label{sec:conclusion}
In this work we have measured 2D ellipticities of galaxies in the CAMELS simulations. We describe how the alignment signals were measured with respect to the dark matter density field, yielding the projected correlation functions $w_{m+}$ (Section \ref{sec:correlation functions}). We validate our results against systematics by ensuring cross correlation functions $w_{m\times}$ are consistent with null and both $w_{m+}$ and $w_{m\times}$ are consistent with null under random ellipticity rotations.

One of the main results of this work is the demonstration that the intrinsic galaxy alignments depend on both cosmological and astrophysical parameters (Section~\ref{sec:results}). By taking advantage of the CAMELS suite's variation of parameters (Table~\ref{tab:camels_params}), we measure how alignment signals respond to changes in these parameters. Our analysis reveals robust correlations between the alignment amplitude and key parameters: $\Omega_m$, $\sigma_8^2$ and the supernova feedback amplitudes $\log A_{\text{SN1}}$ and $\log A_{\text{SN2}}$.

Motivated by standard theoretical models of intrinsic alignment (Section~\ref{sec:Intrinsic alignment theory}) we (partially) removed the density field dependence of $w_{m+}$ by dividing by the density auto-correlation $w_{mm}$, giving the ratio $\frac{w_{m+}}{w_{mm}}$. In Section \ref{sec:wm+wmm}, we find that the correlation with $\sigma_8^2$ is indeed reduced in this ratio, consistent with expectations from the NLA and TATT models. However, the dependence on supernova feedback parameters remains significant, suggesting that astrophysical processes continue to influence the alignment signal beyond what is captured by the matter density field alone.

We further investigated the role of intrinsic ellipticity on alignments by normalizing galaxy ellipticity magnitudes to construct orientation-only catalogues (Sections \ref{sec:ell norm} \& \ref{sec: ell norm analyse}). Significant Pearson correlations in Table~\ref{tab:camels_params} show that orientation-only samples were still dependent on $\log A_{\text{SN1}}$ and $\log A_{\text{SN2}}$. This demonstrates that supernova feedback influences not only the ellipticity magnitude, but also the orientation of galaxy shapes relative to the surrounding dark matter distribution.

These simulations are too small to show significant AGN feedback, so the lack of a significant AGN impact on the alignment signal is to be expected. Given the impact of supernova feeback, we can expect that AGN will also have an effect on the intrinsic alignment signal.  

In separating galaxy catalogues into two samples split by sSFR (at $10^{-10.5}$ /yr), we reveal that star forming galaxies (those above this cut) exhibit alignment amplitudes that are weaker by an order of magnitude compared to quiescent galaxies. This supports the established view that galaxy morphology and star formation activity modulate intrinsic alignments \citep{hirata2007, mandelbaum2011, 2011A&A...527A..26J}.

Section~\ref{sec:SFR and alignments}, examines parameter correlations within the sSFR-split samples. We found that the alignment amplitudes of star-forming galaxies are largely insensitive to variations in $\sigma_8$, the amplitude of matter density fluctuations.

This work has also enabled Figure~\ref{fig:quiescent_prop}, which reveals that the fraction of quiescent galaxies increases with higher $\sigma_8$. Since quiescent galaxies exhibit alignments approximately an order of magnitude larger than those of star-forming galaxies, their growing presence enhances the overall alignment signal in the full population. A central insight here is that $w_{m+}$ inherits an additional $\sigma_8$ dependence, beyond the $\sigma_8^2$ scaling from the density field that is expected from linear approximations described in Section~\ref{sec:Intrinsic alignment theory}.

Intriguingly, if we assume the NLA or TATT models to be true, this additional $\sigma_8$ dependence appears to contradict our earlier finding in Figure~\ref{fig:dep_wdp/wdd}, which shows that the ratio $\frac{w_{m+}}{w_{mm}}$ removes the linear $\sigma_8^2$ dependence in $w_{m+}$ associated with the density field. The observed cancellation, together with the trend shown in Figure~\ref{fig:quiescent_prop}, suggests that the ratio may also remove the additional $\sigma_8$ dependence introduced by the increasing fraction of quiescent galaxies. We leave this investigation to future work.

{The final part of Section~\ref{sec:SFR and alignments} and Table~\ref{table:pearson_vals} reveals that supernova feedback has a strong impact on the intrinsic alignments of star-forming galaxies: the $w_{m+}$ correlation is more sensitive to variations in the baryonic feedback parameters $A_{\text{SN}1}$ and $A_{\text{SN}2}$ than to changes in $\sigma_8$.}

{Moreover, while $A_{\text{SN}1}$ primarily modulates the intrinsic alignments of star-forming galaxies, in the quiescent population it instead regulates their specific star-formation rates (Figure~\ref{fig:ssfr e ASN1}). This dichotomy suggests that $A_{\text{SN}1}$ drives two distinct mechanisms that imprint on the correlation function in high and low-SFR galaxies. Details about these mechanisms remain to be uncovered and provide opportunities for further work.}

{Together, our results show that the intrinsic alignments reflect a complex interplay between cosmology and astrophysical feedback, an understanding of which is a necessary step to better model alignments in the next generation of galaxy surveys.}

\section*{Acknowledgements}

We thank Lorne Whiteway and Benjamin Joachimi for their helpful comments and discussions on the manuscript (in particular the former's cromulent suggestion regarding Simpson's paradox). We also thank Christopher Lovell for a helpful discussion during his visit to UCL. NJ is supported by the ERC-selected UKRI Frontier Research Grant EP/Y03015X/1 and by STFC Consolidated Grant ST/V000780/1. DB was supported by the STFC UCL Centre for Doctoral Training in Data Intensive Science (grant ST/W00674X/1) including departmental and industry contributions.

\section*{Data Availability}
Github repository: \url{https://github.com/danielbilsborrow6/CAMELS-correlations} contains code for all results in this paper.
Data for positions and projected ellipticities of galaxies in the LH set of CAMELS IllustrisTNG is also available.



\bibliographystyle{mnras}
\bibliography{article} 




\appendix

\section{Cosmology dependence of the ratio under NLA} \label{sec:ratio_dependence}

We confirm that, under the NLA model, the ratio $\frac{w_{m+}}{w_{mm}}$ has relatively little dependence on $\sigma_8$ and $\Omega_m$.  Following Eq.~\ref{eq:wm+_pk} and Eq.~\ref{eq:wmm_pk}, the ratio is given by:
\begin{equation}\label{eq:w_ratio}
\begin{split}
\frac{\bar{w}_{m+}}{\bar{w}_{mm}}(r_\perp) &=  -A_{IA} \,c_1\,\rho_{\rm crit}\Omega_m \times \\ & \frac{\int_0^\infty \frac{\mathrm{d}k_z  }{\pi^2} \int_0^\infty \mathrm{d}k_\perp \, \frac{k_\perp}{k_z} \, P_{\delta}(k)  \sin(k_z \Pi_{\mathrm{max}}) \, J_2(k_\perp r_\perp)}{\int_0^\infty \frac{\mathrm{d}k_z  }{\pi^2} \int_0^\infty \mathrm{d}k_\perp \, \frac{k_\perp}{k_z} \, P_{\delta}(k)  \sin(k_z \Pi_{\mathrm{max}}) \, J_0(k_\perp r_\perp)} \ .
\end{split}
\end{equation}
It is not generally possible to simplify this further for a general $P_\delta$. However, $\sigma_8$ defines the overall amplitude for the  \textit{linear} power spectrum $P_{\delta, \rm lin}(k)$. As $\sigma_8^2 \propto P_{\delta, \rm lin}$, $\sigma_8$ is cancelled in the ratio above. So, in the pure linear model, the insensitivity to $\sigma_8$ is exact. 

More generally, we can demonstrate the insensitivity of the theoretical ratio $\frac{\bar{w}_{m+}}{\bar{w}_{mm}}$ via numerical evaluation of Eq.~\ref{eq:w_ratio}. We use the \texttt{CCL} package~\citep{2019ApJS..242....2C} to calculate the theoretical spectra $P_{\delta I}(k)$ and $P_{\delta}(k)$ under the NLA model. The conclusions are not affected by choice of non-linear model \citep[e.g. HALOFIT:][]{2003MNRAS.341.1311S, 2012MNRAS.420.2551B}

\begin{figure}
	\includegraphics[width=\columnwidth]{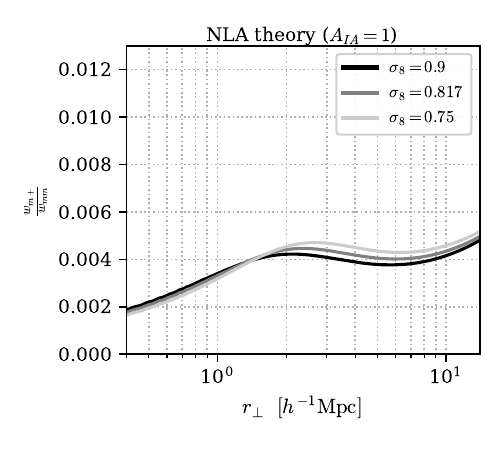}
    \caption{Theoretical ratio of the density–shape $w_{m+}$ correlation functions to density–density correlation function $w_{mm}$ for varying values of $\sigma_8$ assuming the NLA model (at redshift $z=0$). The effect of $\sigma_8$ on the ratio $w_{m+}/w_{mm}$ is $\sim 4$ per cent.}
    \label{fig:wdp/ratio_theory_sigma8}
\end{figure}

Figure~\ref{fig:wdp/ratio_theory_sigma8} shows $\frac{\bar{w}_{m+}}{\bar{w}_{mm}}$ under the NLA model for three $\sigma_8$ values. All other cosmological parameters are kept fixed: $\Omega_m = 0.3$, $\Omega_b=0.046$, $h=0.7$,  $\sigma_8=0.817$, $n_s=0.9646$. As expected, the result is only mildly affected by the value of $\sigma_8$.

\begin{figure}
	\includegraphics[width=\columnwidth]{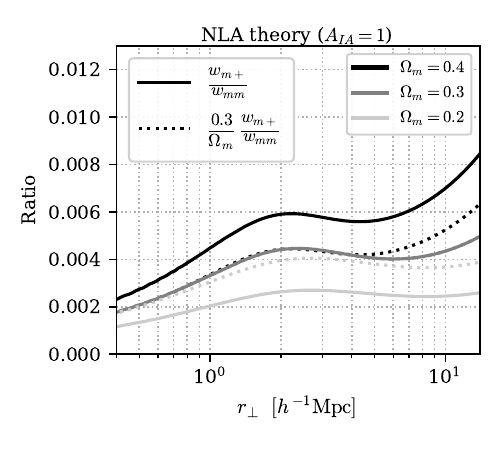}
    \caption{Theoretical ratio of the density–shape $w_{m+}$ correlation functions to density–density correlation function $w_{mm}$ for varying values of $\Omega_m$ assuming the NLA model (at redshift $z=0$). By dividing by $\Omega_m$, the differences between the theoretical ratios are significantly reduced.}
    \label{fig:wdp/ratio_theory}
\end{figure}

Figure~\ref{fig:wdp/ratio_theory} shows $\frac{1}{\Omega_m}\frac{\bar{w}_{m+}}{\bar{w}_{mm}}$ under the NLA model for three $\Omega_m$ values. The prefactor $\frac{1}{\Omega_m}$ removes the explicit $\Omega_m$ dependence in the NLA alignment amplitude. The result is only mildly affected by the value of $\Omega_m$. Some dependence is expected, as $\Omega_m$ changes the average slope of the matter power spectrum $P_\delta(k)$, which does not exactly cancel due to the different Bessel functions $J_0$ and $J_2$. {We reiterate that for the ranges we consider in the CAMELS simulations the pure-linear model assumption described above is not exact and $\frac{w_{m+}}{w_{mm}}$ certainly has non-linear contributions. For this reason and others, the rest of the paper takes on a model-agnostic viewpoint.}

\section{Density autocorrelation integral constraint correction} \label{sec:integral_constraint_correction}

To fit the per-simulation correction to the integral constraint bias, we model the $w_{gg}$ as power law:
\begin{equation}
w_{gg, \mathrm{th}}(r_\perp) = A \left(\frac{r_\perp}{r_0}\right)^{-\gamma} - C
\end{equation}
This form is standard ~\citep[e.g.][]{Peebles:1980yev}, but is still approximate at small scales. However, as we discuss below, our conclusions are not, in fact, sensitive to the exact value of $C$. 

Assuming a diagonal covariance matrix $\Sigma$ with diagonal entries given by the per bin variance from all measurements. This is an overestimate of the variance, but the parameter fit is dependent on relative variance between the per $r_\perp$ bins, which shows the expected heteroscedasticity with largest variance at large scales (Figure~\ref{fig:mean_corr1}).

\begin{figure}
	\includegraphics[width=\columnwidth]{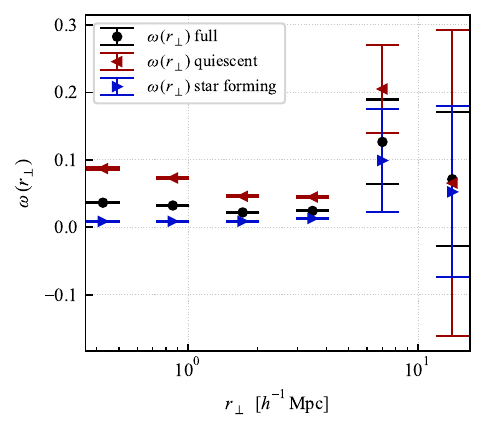}
    \caption{Ratio of the density–shape to density–density correlation functions, averaged across all catalogues: $\bar R_{j}(r_{\perp,j})$. Unlike Figure~\ref{fig:wdp/wdd}, this mean ratio is calculated using varying $C$ values instead of a constant where $C$ is calculated using linear interpolation. Mean ratios are shown for the full, quiescent, and star-forming galaxy samples.}
    \label{fig:wdp/wdd+C}
\end{figure}

We find the values of $A$, $\gamma$, $r_0$, and $C$ that maximize the likelihood:
\begin{equation}
\ln \mathcal{L}
= -\frac{1}{2}
\sum_{i=1}^{N}
\left[
\frac{\bigl(w_{gg,\mathrm{obs}}(r_{\perp \ i}) \;-\; w_{gg,\mathrm{th}}(r_{\perp\ i})\bigr)^2}{\Sigma_{i,i}}
\;+\;
\ln\!\bigl(2\pi\,\Sigma_{i,i}\bigr)
\right].
\end{equation}
This gives us a value of $C$ per simulation. The ratio $R^{(i)}_j$ (Eq.~\ref{eq:ratio}) is recalculated with these varying integral constraint $C$ values. The resulting mean ratio across catalogues $\bar R(r_{\perp,j})$ is displayed in Figure~\ref{fig:wdp/wdd+C}.

Our conclusions are unchanged when using varying per-simulation values of $C$ vs a constant $C=1$. Our conclusions (Figure~\ref{fig:dep_wdp/wdd}) are also unchanged if we take a naive $C=0$, even though the ratio $w_{g+}/w_{gg}$ becomes negative, because the scales affected by the integral constraint additive bias are also those with the largest variance (due to fewer modes at large scales).

\bsp	
\label{lastpage}
\end{document}